\documentclass[%
reprint,
showpacs,
preprintnumbers,
bibnotes, 
amsmath,
amssymb,
aps,
prl,
floatfix,
]{revtex4-2} 

\usepackage{natbib}
\usepackage{graphicx}
\usepackage{amsmath,amsthm}
\usepackage{amsfonts}
\usepackage{amssymb}
\usepackage{bbold}
\usepackage{times}
\usepackage{mathrsfs}
\usepackage{color}
\usepackage{gensymb}
\usepackage{times}
\usepackage{bm}
\usepackage{stmaryrd}
\usepackage{amsmath,amssymb,amsthm,amsfonts,eucal}
\usepackage[english]{babel}
\usepackage{mathtools, nccmath}
\usepackage{epsfig,bm}
\usepackage{tabularx}
\usepackage{bm}
\usepackage{wrapfig}
\usepackage{tikz}
\usepackage{latexsym}
\usepackage{esint}
\usepackage[normalem]{ulem}
\usepackage{epstopdf} 
\usepackage{accents}
\usepackage[makeroom]{cancel}
\usepackage[bbgreekl]{mathbbol}
\usepackage[utf8]{inputenc} 

\DeclareSymbolFontAlphabet{\mathbb}{AMSb}
\DeclareSymbolFontAlphabet{\mathbbl}{bbold}

\renewcommand\d\delta
\newcommand\D\Delta

\newcommand\cD{\mathcal{D}}
\newcommand\cE{\mathcal{E}}

\newcommand\cJ{\mathcal{J}}

\newcommand\s{\sigma}

\newcommand\beq{\begin{equation}}
\newcommand\beqn{\begin{eqnarray}}
\newcommand\eeqn{\end{eqnarray}}
\newcommand\eeq{\end{equation}}

\newcommand{\jump}[1]{\left\llbracket #1 \right\rrbracket}
\usepackage{bm} 
\newcommand{\curly}[1]{{\left\{ #1 \right\}}}

\begin{document}

\title{Active chemo-mechanical solitons} 
\author{L. Truskinovsky}
\affiliation{\it  PMMH, CNRS -- UMR 7636, CNRS, ESPCI Paris, PSL Research University, 10 rue Vauquelin, 75005 Paris, France}
\author{G. Zurlo}
\affiliation{\it  School of Mathematical and Statistical Sciences, University of Galway, University Road, Galway, Ireland}

\begin{abstract}
In many  biological systems  localized mechanical information is transmitted by  mechanically    neutral chemical signals. Typical examples include  contraction waves in acto-myosin cortex  at cellular scale and peristaltic waves at  tissue level. In such systems, chemical activity is transformed into mechanical deformation by distributed motor-type mechanisms represented by continuum degrees of freedom. To elucidate the underlying principles of chemo-mechanical coupling, we here present the simplest example, involving directional motion of a localized solitary wave in a distributed mechanical system, guided by a purely chemical cue. Our main result is that   mechanical signals can be   driven by chemical activity in a highly  efficient manner.
\end{abstract}

\maketitle

\section{Introduction}

While  molecular biology and biochemistry  clearly play dominant roles in   biological processes,    mechanics has recently  emerged as another  crucial co-player \cite{savin2011growth,coen2023mechanics,
lappalainen2022biochemical,truskinovsky2014mechanical}.  However,   how chemically generated microscopic forces are transformed into a functional mechanical deformation at the macroscale still remains opaque \cite{slater2023emergence,wang2024collective}. 

Of particular interest  in this respect are the mechanisms of directional transmission of localized mechanical information by mechanically-neutral chemical signals. For instance,  cardiac muscles have the ability to actively generate functional dynamic contractions \cite{bagshaw1993muscle}, fluid propulsion of many cellular organisms is achieved by deformation waves  \cite{lindemann2024mechanics,beeby2020propulsive}, fronts of cytoskeletal reinforcement were observed during epithelial monolayer expansion   \cite{serra2012mechanical,boocock2021theory} and finally  
peristalsis in digestive tract   performs the most remarkable directed  mechanical activity \cite{sinnott2017peristaltic,quillin1999kinematic,gorbushin2021peristalsis}. 

It is known that the inner working of the implied chemical machinery can be modeled by various excitable reaction-diffusion systems \cite{pismen2006patterns}. However, it remains unclear how a fueling chemical signal can transform into a propagating  localized wave-like deformation field, and how such chemo-mechanical system can attain the observed high levels of energetic efficiency \cite{perez2024excitable, kindberg2020forced,alisafaei2021long,dean2023mechanics}. 

To elucidate the underlying principles of chemo-mechanical coupling we present in this paper   the simplest example of a directional motion of a  solitary wave  in a distributed mechanical system driven by a  mechanically neutral signal of chemical origin. 

We build on a previous work where   other   systems   displaying active chemo-mechanical coupling  have been extensively studied, see for instance,  \cite{livne2024self,levin2020self,manna2022self,duffy2025programming,li2021chemically,sarkar2025mechanochemical,zakharov2021mechanochemical,cohen2024locomotion}. In particular, various  models have been proposed showing  that  elastic instabilities  can be produced amd dynamically regulated  by  excitable reaction-diffusion systems  \cite{yin2022three,feng2025mechano,ihuaenyi2025learning,picardo2025active,li2021chemically,xiong2025chemo} 

In this paper we are concerned with a particular aspect of such dynamic actuation. Specifically, we show  that  chemo-mechanical  driving of localized solitary waves (solitons, for simplicity) can be  mechanically efficient in the sense that  the mechanical energy needed to activate the pulse at the leading  edge can be transported from trailing edge where it is released due to the the  pulse  closure. As we show,  such   self-sustaining machinery can indeed be operative   due to the existence of a  mechanism of an internal  energy transport through wave dispersion \cite{kartashov2011solitons,truskinovsky2014solitary,  ming2018solitons}. 
In this case   mechanical signal can   propagate without taking any energy from the triggering chemical signal due to an on board energy-harvesting. 

To highlight ideas we consider in this exploratory paper only a purely mechanical side of the corresponding energy transport mechanism without invoking in any detail the underlying chemical machinery \cite{dudchenko2012self,ambrosi2012active,brunello2024regulating,caruel2018physics}. Specifically, we restore to the simplest phenomenological approach in the description of active stress generation ~\cite{prost2015active,kruse2005generic}. Using such stylized framework, we show that a mechanical solitary wave can be indeed generated chemically even in a \emph{linear} mechanical system. Moreover, we show that the chemo-mechanical coupling in such model is maximally efficient, which means that the active chemical system does not need to expend energy to sustain the mechanical signal beyond what is needed to maintain chemical activation.
 
A realistic example  of the implied chemo-mechanical coupling can be found, for instance,  in \cite{Levin2020PRL} where the authors present an experimental realization of the  propagating reaction-diffusion fronts generated by Belousov-Zhabotinsky  reaction which can induce localized deformation in a  gel.  Similarly,  in \cite{Livne2024PNAS} the authors   recreate  in vitro the actomyosin cytoskeleton and show that some of its  buckling type  shape deformations  can be driven actively by  myosin motors. Vis-a vis our schematic model, the propagation of motor-induced activity  can be viewed as the analog of our  chemical signal which also generates  pre-stress responsible for dynamic buckling.  More elaborate chemo-elastic interactions are discussed, for instance,   in \cite{Li2021PNAS} and in such  perspective,  our minimal formulation is intentionally schematic.  However,  due to its full analytical tractability, the proposed model  allowed us to reveal  a novel  fundamental mechanism of dispersive energy transport expected to be operative in such active systems. 
 
Our main technical tool is the theory of  dispersive systems   allowing    dynamical energy transport \cite{whitham2011linear}. We build on the   ideas developed  in the theory of  configurational defects in crystal lattices  \cite{kresse2004lattice,slepyan2012models}, which can produce lattice scale energy radiation but  can be also  dynamically guided by lattice waves  \cite{gorbushin2020frictionless,slepyan2012models}. The main goal of this study   is to show that   similar dispersion-induced mechanisms  can ensure that the energy generated by chemically activated sources  can be fully transmitted to chemically supported sinks which  would then make  a loss-free propagation of mechanical information possible.

\section{The model}

\begin{figure}[h!]
\centering
\includegraphics[width=0.6\columnwidth]{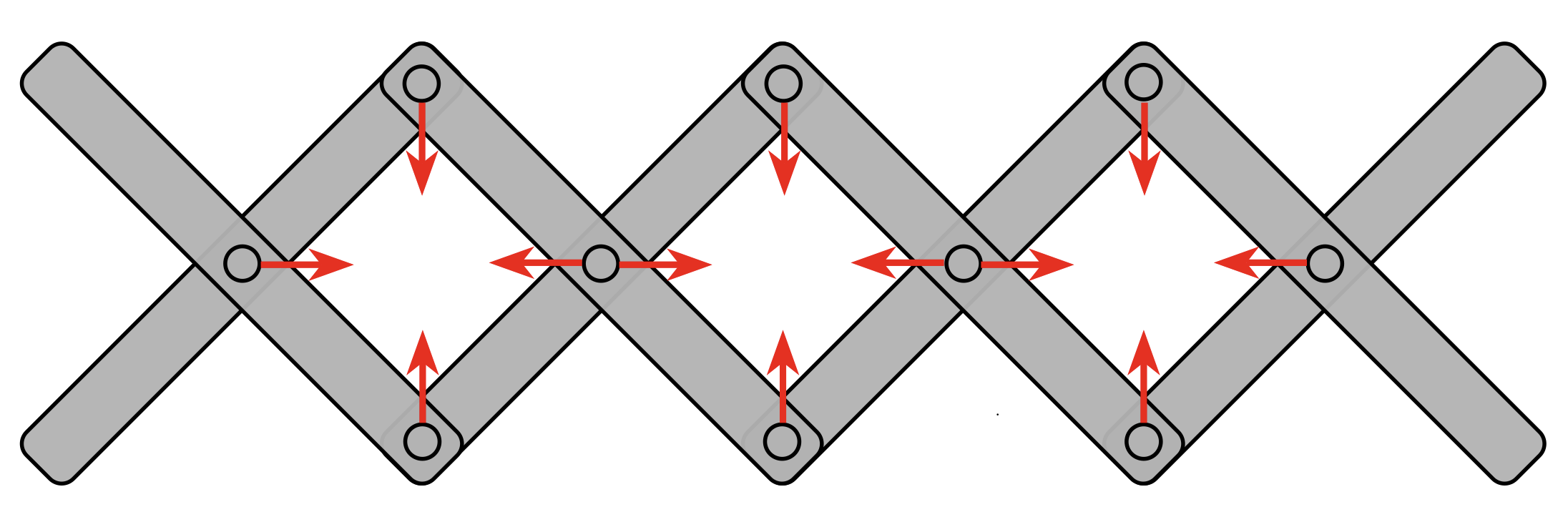}
\caption{\label{fig:12}\footnotesize{Mechanical model of pantograph with integrated active pulling (contractile) elements represented by red arrows. While the underconstrained pantograph structure is not rigid and contains a longitudinal soft mode, the presence of active elements makes it overconstrained.}}
\end{figure}

In the interest of analytical transparency, we consider only a 1D dynamic problem with  $u(x,t)$ denoting  a scalar displacement field (longitudinal or transverse), $x$ being a reference coordinate.    Having in mind the elastic softness of the modeled biological prototypes, we assume that the system is elastically degenerate (infinitely stretchable), which brings into the theory a continuous symmetry.
The corresponding Goldstone (soft) mode is described by the strain variable $u_{x}$, where the subscript indicates partial derivative. When the strain is a slowly varying inhomogeneous field, the corresponding small energy cost can be captured by a quadratic function of strain gradients, meaning that the implied infinitely soft material still carries gradient (bending) rigidity. 
The corresponding free energy is then 
$$
\int   \frac{\kappa}{2} u_{xx}^2 dx
$$ 
where $\kappa$ is the bending modulus,   and subscripts denote partial derivatives, e.g., $u_x \equiv \partial u / \partial x$, $u_{xx} \equiv \partial^2 u / \partial x^2$, etc.  Such energy emerges, for instance, in continuum representation of a deliberately minimalistic pantographic structure, as shown in Fig.\ref{fig:12}. It is built of inextensible but flexible beams, connected through ideal pivots. While the  structure is floppy, showing zero stiffness  when it is deformed uniformly\cite{schenk2014zero}, inhomogeneous deformations are energetically penalized due to nonzero bending rigidity of the beams \cite{alibert2003truss}. More complex examples of bending-dominated structures can be found in the theory of high contrast elastic composites \cite{boutin2013experimental,camar2003determination}; in such systems the  second order elasticity appears already at the leading order in the homogenization (continuum) limit, which justifies our choice of the elastic energy.

In Fig. \ref{fig:12} we also show that the under-constrained pantograph structure can be stabilized through the pre-stress by active elements which eliminate the corresponding soft (floppy) modes.  To effectively incorporate such elements into the model, we need to account for a term in the energy of the form  $$\int \sigma_a (x,t) u_x dx.$$    
We further assume that the (pre)stress  $\sigma_a$ is piecewise constant spin variable taking three values $ \lbrace \sigma_0,0,-\sigma_0 \rbrace$, intended to represent active stretching, deactivation and active compression, accordingly. In Fig.\ref{fig:12} we illustrate the fact that all of these states can be achieved by using antagonistic, purely contractile elements;  it also shows that contraction can be amplified by the   arrangement  which  turns small microscopic deformations into large macroscopic 
movements \cite{mahadevan2000motility}.
  
To allow for dispersive non-dissipative energy transport in the model  we  also need to account for (pseudo)inertia by introducing  into the energy the term 
 $$
\int \frac{\rho}{2}\,u_t^2\, dx.
 $$ 
The interpretation of  such term as the actual kinetic energy may still make sense due to our choice of ultra-soft stretching elasticity, which turns into supersonic even for extremely slow mechanical signals.  However, in physiological setting, the apparent mass density $\rho$ should be rather viewed as bringing into the constitutive model an internal time scale, describing the functional, activity-induced, time delay. Such delay was found relevant, for instance, in the description of the activity of molecular motors, stretch receptors, ERK waves, etc. \cite{badoual2002bidirectional,serra2012mechanical,boocock2021theory}.

\section{Variational formulation}  


In view of the conservative nature of the mechanical part of the ensuing model,  it will be convenient to obtain the corresponding dynamic equations  and boundary conditions using  the variational  Hamilton (action) principle. The  Lagrange  function  in our case has the  form   
\begin{equation} 
L= \frac{1}{2}\rho  u_t^2  -\sigma_a u_{x}  -\frac{1}{2} \kappa u_{xx}^2
\end{equation} 
where the inhomogeneous controls $ \sigma_a(x,t)$ are assumed to be given. It will be also convenient to first rewrite the problem in more general terms by introducing the variables $q^1=x$ and $q^2=t$, and presenting the displacement field in the form $u(q^{\alpha})$, with Greek indices $\alpha,\beta,\gamma = 1,2$ in the remainder of this section. The corresponding action functional  is
 \begin{equation}\label{contu}
 {\cal{L}} =\int_{\Omega} L(q^\alpha, u_{\alpha}\, , \, u_{\alpha\beta})\, dq^1 dq^2
 \end{equation}
where we introduced a two-dimensional \emph{space-time} domain $\Omega$ describing the evolution of a body between two time instants \cite{gavrilyuk2020stationary},   and where again $u_{\alpha}=\partial u/\partial q^{\alpha}$. The corresponding  Euler-Lagrange equations, representing the balance of linear momentum,  take the form
$$
\frac{\partial}{\partial q^{\alpha}}\left(\frac{\delta L}{\delta u_{\alpha} }\right) = 0
$$
where 
$$
\frac{\delta L}{\delta u_{\alpha}} = \frac{\partial L}{\partial u_{\alpha}}-\frac{\partial}{\partial q^{\beta}}\left(\frac{\partial L}{\partial u_{\alpha\beta}}\right)
$$ 
is the variational derivative and the summation over repeated indexes is implied. In our special case we obtain the linear equation 
$$
\rho u_{tt}= \sigma_x
$$
where   
$$
\sigma = \sigma_a    -\kappa u_{xxx}.
$$ 
Suppose next that the domain $\Omega$ contains a propagating surface $\Sigma$ where the function $\sigma_a(x,t)$ suffers a jump discontinuity, while it remains constant outside. On $\Sigma$ the dependence of $L$ on $q^{\alpha}$ is discontinuous, which leads to the possibility of energy sources or sinks. It is natural to assume that the particle trajectories on $\Sigma$ are continuous, so that 
$$
\jump{u}=0
$$ 
where 
$$
\jump{f} =f^{+}-f^{-}
$$ 
with the superscripts $\pm$ denoting the limiting values of $f$. To ensure finiteness of the energy it is also natural to assume that the particle trajectories are smooth, 
$$
\jump{u_x }=0.
$$  
On $\Sigma$ the stationarity of the action  ${\mathcal{L}}$ imposes the conditions representing continuity of   linear momentum  and hypertractions (moments) on the jump \cite{mindlin1962effects,toupin1964theories}: 
\begin{equation}\label{contu1}
\jump{\frac{\delta L}{\delta u_{\alpha}}}n_{\alpha}  = 0
\end{equation}
and 
\begin{equation}\label{contu2}
\jump{\frac{\delta L}{\delta u_{\alpha\beta}}}n_{\alpha} n_{\beta} = 0
\end{equation}
where $n_{\alpha}$ is the  unit vector normal to $\Sigma$ facing the $+$ direction. Note also that mass balance on $\Sigma$ is guaranteed by the kinematic compatibility condition  
$$
\jump{u_{\alpha} }= \mu n_{\alpha}
$$ 
where  $\mu$ is a scalar.  The spatial $n_x$ and temporal $n_t$ components of the normal vector  to $\Sigma$ are related through 
$$
n_t=-n_{x} D
$$ where $D$ is the normal velocity of the discontinuity, and we can always assume that $n_x=1$.  Therefore, the jump conditions (\ref{contu1},\ref{contu2}) reduce  to 
\begin{equation}\label{contu1bis}
\jump{\sigma+\varrho D u_t} = 0
\end{equation}
and  
\begin{equation}\label{contu2bis}
\jump{m} = 0.
\end{equation}
respectively, where we introduced a special  notation for the moments (hypertractions) 
\begin{equation}
m=\kappa u_{xx}.
\end{equation} 
 Furthermore, eliminating $\mu$ we obtain the relation 
$$
D\jump{u_{x} }+\jump{u_{t} }=0
$$ requiring that on $\Sigma$ not only the strain $u_{x}$, but also the velocity field $u_{t}$ must be continuous.

As we have already mentioned, on a surfaces of discontinuity  $\Sigma$,   one can expect  sources or sinks of energy. To find the rate of energy release (absorption) on such jump discontinuities, we need to assess the singular behavior on $\Sigma$ of the energy momentum tensor \cite{eshelby1975elastic,truskinovskii1987dynamics}
\begin{equation}
\tau^{\alpha}_{\beta}=L \, \delta^{\alpha}_{\beta}  -u_{\beta}\, \frac{\delta L}{\delta u_{\alpha}}-u_{\gamma\beta}\, \frac{\partial L}{\partial u_{\gamma\alpha}}
\end{equation}
where $\delta^{\alpha}_{\beta}$ is the Kronecker symbol. Specifically, the temporal component of $\tau^{\alpha}_{\beta}$ controls  the  energy balance.
Thus if we can  use Noether identity   
\beq
\frac{\partial \tau^{\alpha}_{\beta}}{\partial q^{\alpha}}=0.
 \eeq
 and then take $\beta=2$ we obtain the equation  $$w_t + \mathcal{J}_x=0,$$ where  $$w= \frac{1}{2}\rho u_t^2   +\sigma_a u_x+  \frac{1}{2} \kappa u_{xx}^2$$ is the  energy density and $$\mathcal{J}=-\sigma u_t-\kappa u_{xt} u_{xx}$$ is the flux. To compute   a singular   energy release (absorption) on the surface of discontinuity $\Sigma$ we need to compute the normal projection 
 \begin{equation}
\cD=\jump{ \tau^{\alpha}_2}n_{\alpha}.
\end{equation} 
In view of the imposed  continuity conditions  the rate of dissipation at a jump 
is  then
\begin{equation} \label{didss1}
\cD= D \jump{\sigma_a}\curly{u_x}
\end{equation}
where we introduced the average 
$$
\curly{f} =\frac{1}{2} (f^{+}+f^{-}).
$$

\section{Direct formulation}

In view of a somewhat abstract nature of the variational approach, it is instructive to consider in parallel  a  more conventional  continuum mechanical formulation of the same problem which allows one to recover the same relations but in a more pedestrian setting. We won't aim at recovering  the whole classical construction from scratch but only focus on the most nontrivial aspect of the problem related to jump discontinuities.

To be more specific, consider directly   a 1D problem ans set it  in a generic spatial domain $\mathcal{C}=(0,L)$ where we directly define the displacements field $u=u(x,t)$.    

We build our discussion around  the expression for  the rate of change/release of the total energy 
$$
\cE=\int_{\mathcal{C}} w\,dx. 
$$ 
For reasons to become clear later,  we will consider a slightly more general formulation of the model based on  the energy density of the form
$$
 w=\frac{1}{2} {\varrho}  u_t^2+\frac{1}{2}  {E} u_{x}^2+\frac{1}{2}  {\kappa} u_{xx}^2+\sigma_a u_x. 
$$
In addition to the nonclassical bending elasticity term, it now includes also a stretching elasticity term characterized by the energy 
$$
\int\frac{1}{2} E u_{x}^2 dx
$$ 
where  ${E}$ is the stretching  modulus. 

Suppose next that the location of the surface of gradient discontinuity $\Sigma$ is prescribed by the function  $\zeta(t)$ so that  $D=\zeta_t$. We obtain
\beq \label{energy}
\cE_t = \int_{\mathcal{C}}(\varrho u_{tt} - {\hat{\sigma}}_x) u_t\,dx - [{\cJ}]_{\partial\mathcal{C}} - \jump{D w  - {\cJ}}_{\zeta(t)}
\eeq
where now the stress in the system is
$$
{\hat{\sigma}} = \sigma_a + E u_{x} - \kappa u_{xxx}
$$
and the energy flux is defined as
\beq\label{energyflux}
{\cJ} = -\hat{\sigma} u_t - \kappa u_{xx} u_{xt}.
\eeq
We assume for simplicity that  $u=u_x=0$ at the external boundary of the body $\partial\mathcal{C}$ so that $u_t=u_{xt}=0$ as well. Then  $[\cJ]_{\partial\mathcal{C}}=0$ and  there is no outside flux  of energy.

Away from discontinuity, balance of momentum is satisfied
\beq\label{main-equation-motion}
\varrho u_{tt} -  \hat{\sigma}_{x} = 0. 
\eeq 
Note next that  the jump term in \eqref{energy} can be rewritten as 
\beqn\label{jumpgen}
\jump{D w - {\cJ}} 
&=& \varrho D \jump{u_t}\curly{u_t} + D\kappa\curly{u_{xx}}\jump{u_{xx}} \nonumber\\
 &+& D E \curly{u_x}\jump{u_x} + E\jump{u_x}\curly{u_t} +  {\color{black}E\curly{u_x}\jump{u_t}} \nonumber\\
&+& {\color{black}D\jump{\s_a}\curly{u_x}} + {\color{black}D\curly{\s_a}\jump{u_x}} \nonumber\\
&+& \kappa\jump{u_{xx}}\curly{u_{xt}} +  \kappa\jump{u_{xt}}\curly{u_{xx}}\nonumber\\
&+& \jump{{\color{black}\s_a} - \kappa u_{xxx}}{\color{black}\curly{u_t}} + \curly{{\color{black}\s_a} - \kappa u_{xxx}}{\color{black}\jump{u_t}}
\eeqn
 As in our variational analysis we assume that on jump discontinuities the displacement is continuous, $\jump{u}=0$, and therefore $\jump{u_t} + D\jump{u_x} = 0$. Because we have similarly assumed that $\jump{u_x}=0$, we also have  $\jump{u_t}=0$. Furthermore, by time differentiation of  $\jump{u_x}=u_x^+(\zeta(t),t) - u_x^-(\zeta(t),t) = 0$, we obtain that necessarily  
 $$
 \jump{u_{xt} + D u_{xx}} = 0.
 $$
%
 

To imitate our   variational setting we further  assume that the corresponding  tractions $\hat{\sigma}$
and  moments   $m$ are   continuous on the jump discontinuity, so that  $\jump{\hat{\sigma}}=0$ and $\jump{m} =0$. By collecting all the jump conditions, 
\beq\label{jump1}
\jump{u} = 0, \quad \jump{u_{x}}=0, \quad
\jump{u_{xx}} = 0,\quad
\jump{\s_a - \kappa u_{xxx}} = 0. 
\eeq
Finally, when using these jump conditions into \eqref{jumpgen}, we find that the dissipation at every jump set equals
\beq\label{Dgen}
 \mathcal{D} = - \cE_t = D \jump{\s_a} \curly{u_x} 
\eeq 
which expectedly agrees with \eqref{didss1}. Note that if the rate of energy release is computed over an arbitrary region $(a,b)$ that does not contain discontinuities, we obtain  
\beq
\int_a^b (\hat{w}_t + \hat{\mathcal{J}}_x)\,dx = \int_a^b (\varrho u_{tt} - {\hat{\sigma}}_x) u_t\,dx = 0 
\eeq
which after the localization equally expectedly gives  $$\hat{w}_t + \hat{\mathcal{J}}_x=0.$$

\section{Active driving }

Consider next a steadily propagating mechanical solitary wave (pulse)  in the form of a traveling wave  $u(x,t)=u(z)$, where $z=x-Dt$ with $D>0$. We assume that it is driven internally by a chemical signal also in the form of a traveling wave, generating a dynamic active stress distribution $\sigma_a(x,t)=\sigma_a(z)$. 

A reasonable assumption for living (or artificial) systems would be a continuous distribution $\sigma_a$, which would bring an additional length scale of chemo-diffusional nature into the model. However, for analytical simplicity, we here introduce a non-continuous distribution for $\sigma_a$. Specifically, we assume that the chemical activity wave can be represented by a combination of two mutually compensating rectangular pulses of compression and stretching: 
\beq
\sigma_a(z)=
\left\{
\begin{array}{ccc} \label{S}
0 & l<z & \text{(zone 1)}\\
-\sigma_0 & 0 < z < l & \text{(zone 2)}\\
\sigma_0 & -l < z < 0 & \text{(zone 3)}\\
0 & z<-l & \text{(zone 4)}
\end{array}
\right.
\eeq
where $\sigma_0>0$ is the amplitude  of chemical activity  and $2 l $ is the size of the propagating activity  zone, which so far is viewed as a parameter independent on $D$, the velocity of the traveling wave. We must therefore integrate the  resulting linear equation in the four domains  separated by the  discontinuities of $\sigma_a$. 

\begin{figure}[h!]
\centering
\includegraphics[width=0.7\columnwidth]{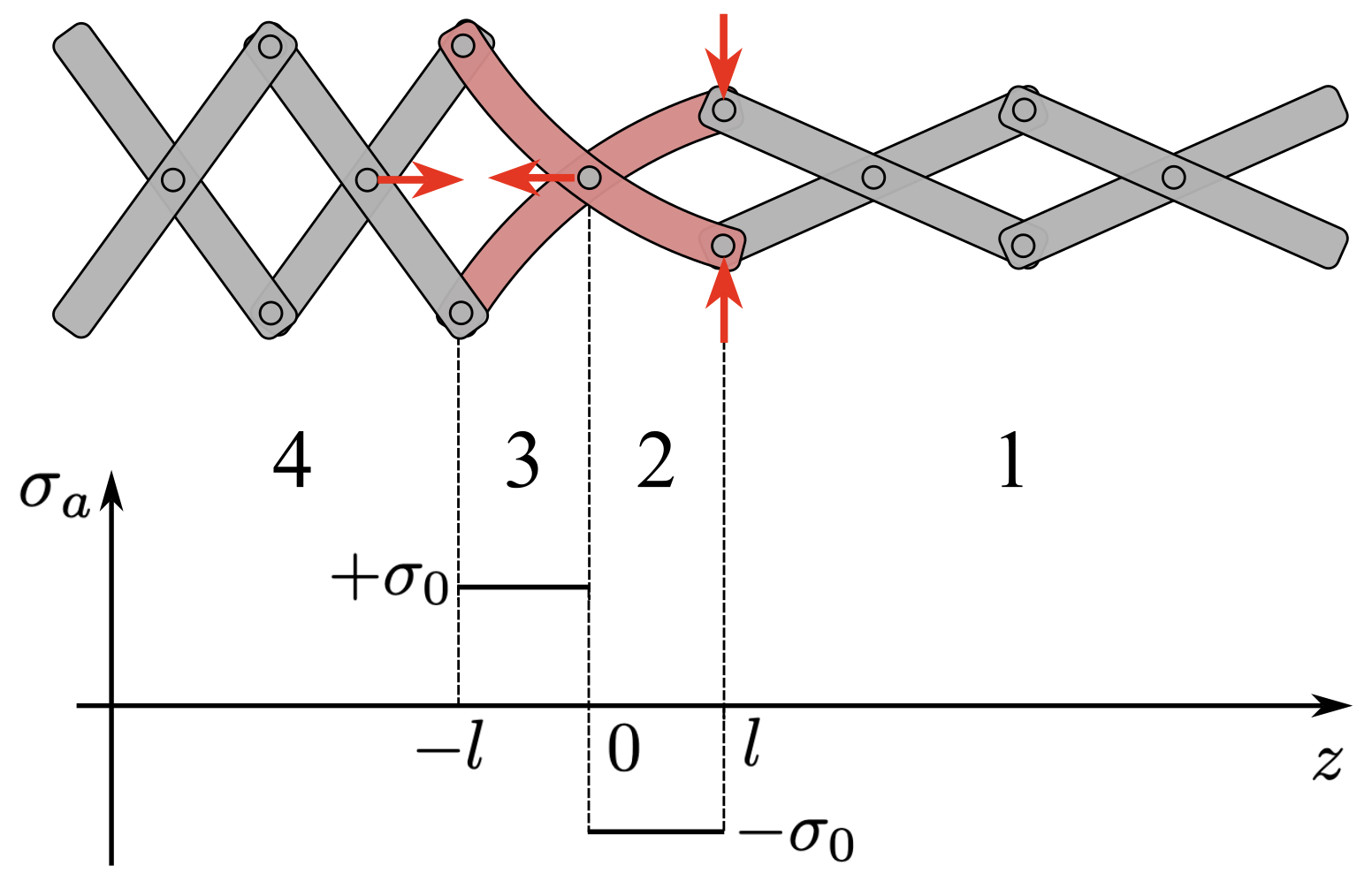}
\caption{\label{activestress}  \footnotesize{Schematic representation of the active loading \eqref{S} in the pantographic structure shown in Fig. \ref{fig:12}. Note that active elements operate only inside the chemically active region.} 
} 
\end{figure}

Because the distribution of active stresses $\sigma_a(z)$ is  piecewise constant, it is  equivalent to a set of localized self-equilibrated  forces (couples) generating active compression  in $0\leq z \leq l$ (zone 2) and  active tension  in $-l\leq z \leq 0$ (zone 2). In Fig.  \ref{activestress} we show schematically the active loading \eqref{S} for the pantographic structure in the case of a traveling wave solution. 

\section{Traveling waves}

Traveling wave solutions are obtained by solving the fourth-order linear ODE 
\begin{equation}\label{main}
E_{\text{eff}} u''(z) - \kappa u''''(z) = -\sigma_a'(z),
\end{equation}
where $u' = du/dz$ and the effective modulus is
$$
E_{\text{eff}} = E - \rho D^2.
$$
Here $E$ is the elastic modulus, $\rho$ the density, and $D$ the wave speed. When $E = 0$, $E_{\text{eff}} = -\rho D^2 \le 0$.  

Interpreting the left-hand side of \eqref{main} as the Euler-Lagrange equation of a static problem with elastic energy
$$
\int \left(\frac{1}{2} \kappa u_{xx}^2 + \frac{1}{2} E_{\text{eff}} u_x^2\right)\,dx
$$
we see that although $E > 0$, inertial effects can make $E_{\text{eff}} < 0$. This mechanism underlies the emergence of a buckling-type instability, that produces different types of traveling waves in the system, as we discuss below.

On the moving surfaces of discontinuity of the active field $\sigma_a(z)$, the solutions to \eqref{main} should be matched by using the corresponding jump conditions. When changing variables, the continuity conditions $\jump{u}=0, \jump{u_x}=0$ on the jumps become 
\beq\label{jumps1}
\jump{u }=0, \qquad \jump{u' }=0
\eeq
whereas the continuity of tractions and hypertractions on the jumps gives  
\beq\label{jumps2}
\jump{\sigma_a-\kappa u''' }=0, \qquad \jump{u'' }=0.
\eeq
As it follows  from \eqref{Dgen} and  \eqref{jumps1}$_1$,   the dissipation at a jump is 
\beq\label{jumps3}
 \mathcal{D} = D \jump{\s_a} u'.
\eeq 
Since $\jump{\s_a}$ differs from zero on such jump, the requirement that the soliton is non-dissipative is equivalent to the condition that  
\beq
u'=0
\eeq
at the  jump. Finally, we observe that according to the general expression \eqref{energyflux}, the flux of energy for the stationary solution is
\beq
\cJ(z) = D\kappa u''^2(z) + Du'(z)\hat\sigma(z). 
\eeq


It will be convenient to work with dimensionless variables  and we set 
$$
l_0=\sqrt{\frac{\kappa}{\sigma_0}}, \qquad
c_0=\sqrt{\frac{\sigma_0}{\rho}}
$$ as the  scales of length and velocity, respectively. The remaining main nondimensional parameters of the problem are  then 
$$
K=\frac{l}{l_0}, \qquad \nu=\sqrt{\frac{E}{\sigma_0}}.
$$
In what follows  we mark dimensionless variables with tildes, 
$$
\tilde z=\frac{z}{l_0},\qquad
\tilde u=\frac{u}{l_0},\qquad
\tilde D = \frac{D}{c_0}.
$$ 
Then, with reference to \eqref{S}, zone 1 corresponds to $K\leq \tilde z$, zone 2 to $0\leq \tilde z\leq K$, zone 3 to $-K\leq \tilde z\leq 0$, and zone 4 to $\tilde z \leq -K$.

The simplicity of our setting allowed us to solve the problem analytically in its full generality. Depending on the values of the nondimensional parameters we identified two main  regimes: supersonic and subsonic.


\section{Supersonic regime}

Suppose first that the chemical driving impulse  propagates with velocity 
\beq
\tilde D \geq  \nu.
\eeq   
It is natural to refer to this regime as supersonic. Note that one is always in this regime when ``stretching'' elasticity is negligible, meaning $E\approx 0$ and therefore effective sonic velocity $\nu$ is small.

In this  regime  the general solution of  \eqref{main} in the regions 1-4 is a combination of linear and periodic functions, 
\beq\label{mainsolution}
u_a(z) = c_{a1} + c_{a2} z + c_{a3}\cos(\beta z/l_0) + c_{a4}\sin(\beta z/l_0)
\eeq
where  $a=1,..,4$. In \eqref{mainsolution}  we introduced the notation 
$$
\beta=\sqrt{\tilde D^2 - \nu^2}. 
$$

\begin{figure}[thb!]  
\centering
    \includegraphics[width=0.6\columnwidth]{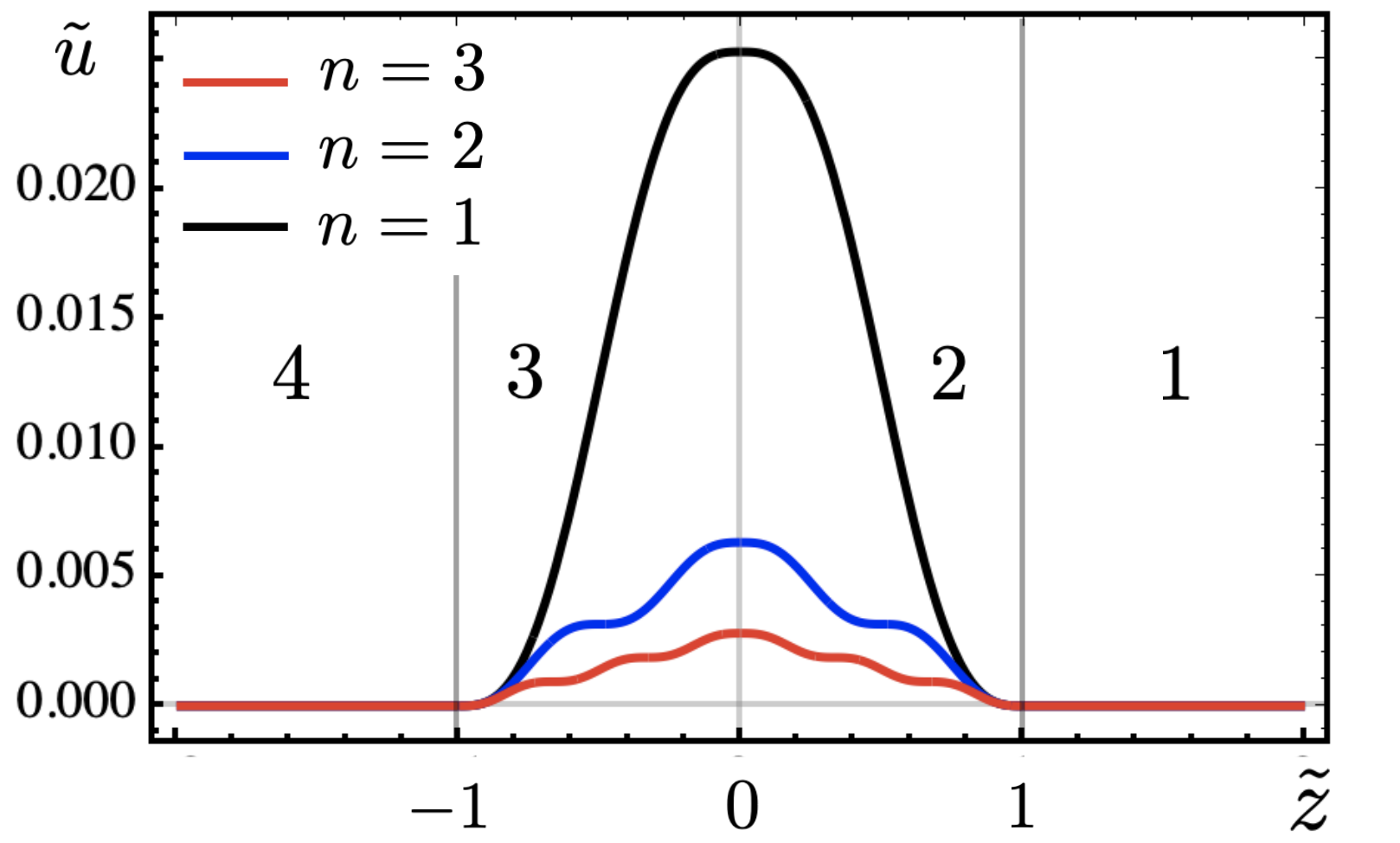} 
    \caption{\footnotesize{Dimensionless displacement field in compactons for different values of $n$. Here $K=1$ and $-2\leq \tilde z \leq 2$.}} \label{shape1}
\end{figure}

Consider first  solutions  with compact support localized in the active region.  To define such regimes  we   set $$\tilde u_1(\tilde  z)=\tilde u_4(\tilde z)=0.$$ Under these constraints, the system  \eqref{jumps1},\eqref{jumps2}, \eqref{jumps3} becomes over-determined and  non-trivial solutions   exist only at the special values  of parameter $\beta$. Specifically, we obtain a discrete spectrum  of admissible values  
\beq\label{betan}
\beta_n = \frac{2 l_0}{l} n \pi, \qquad n=1,2,...
\eeq
The   corresponding  {\it compactons} are of the form
\beq 
\tilde u_{2,3}(\tilde z) = \frac{K^2}{8 n^3 \pi^3} \left(2 n \pi (K \mp \tilde z) \pm K \sin\left(\frac{2 n \pi  \tilde z}{K}\right)\right).
\eeq
The structure of these solutions  is illustrated in  Fig.\ref{shape1}  for  the smallest values of the integer  parameter $n$. 
Note that they  require for their existence  a very special special values  of the velocity of the active signal given by the formulas 
\beq\label{Dn}
D_n= \left({\left(\frac{2 n \pi}{l}\right)^2{\frac{\kappa}{\rho}}+ \frac{E}{\rho}}\right)^{1/2}.
\eeq
Note, in particular,   that $D_n$ does not depend on the level of active prestress $\sigma_0$ and that narrow pulses (small  $l$)    move faster than  broad ones (big  $l$).  More generally, the  presence of a mechanical constraint  on the  spectrum of  admissible velocities of  chemical signals   suggests that there is  a feedback between  mechanical and chemical systems. In this sense, the constructed  compact  pulses are truly ``chemo-mechanical''.
\begin{figure}[thb!]  
\centering
    \includegraphics[width=0.6\columnwidth]{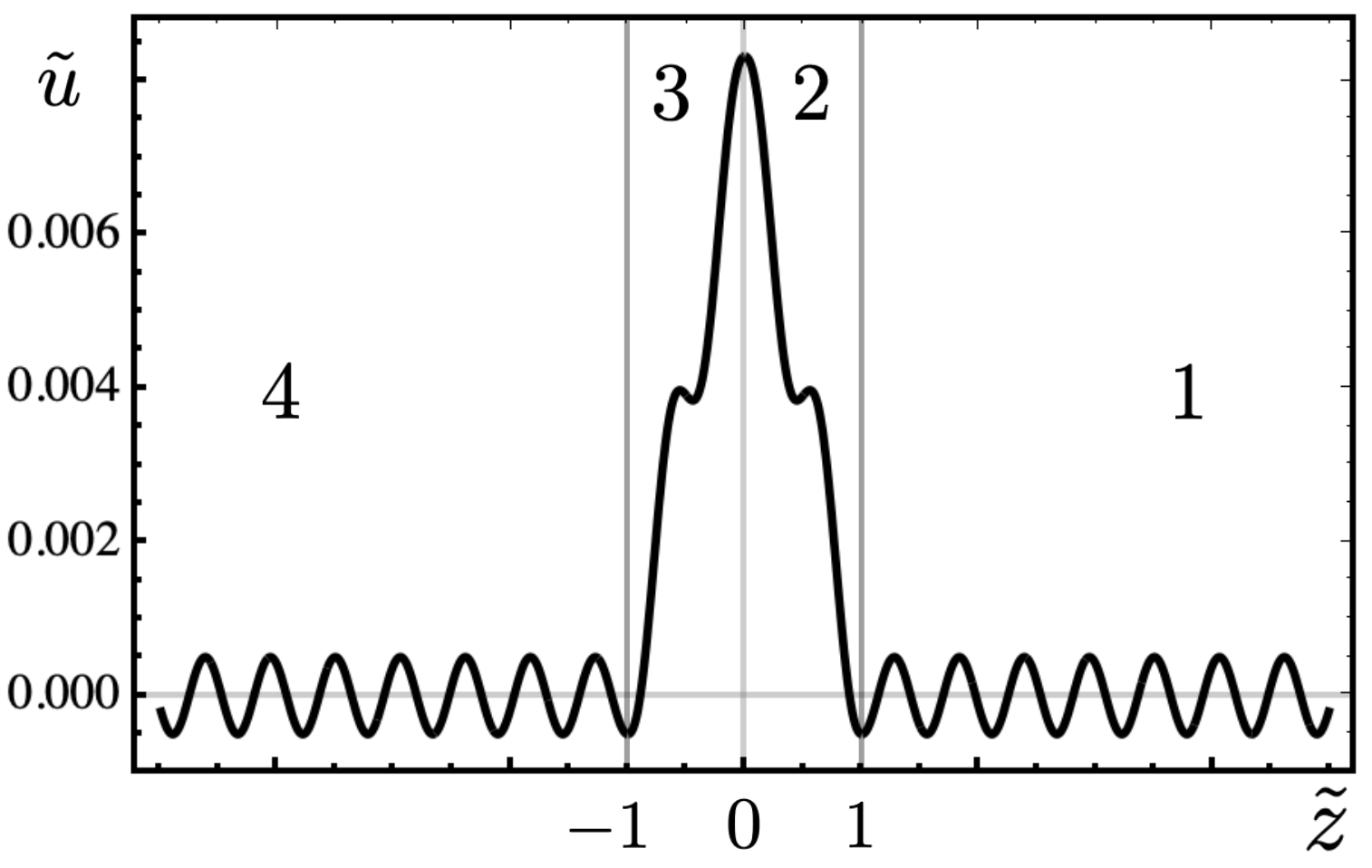} 
    \caption{\label{nano}\footnotesize{Dimensionless displacement field in ({\it  nanopterons}) obtained  for $\beta=0.9 \beta_2$. Here $K=1$ and $-5\leq \tilde z \leq 5$.}} 
\end{figure}

\begin{figure}[h!]  
\centering
    \includegraphics[width=0.7\columnwidth]{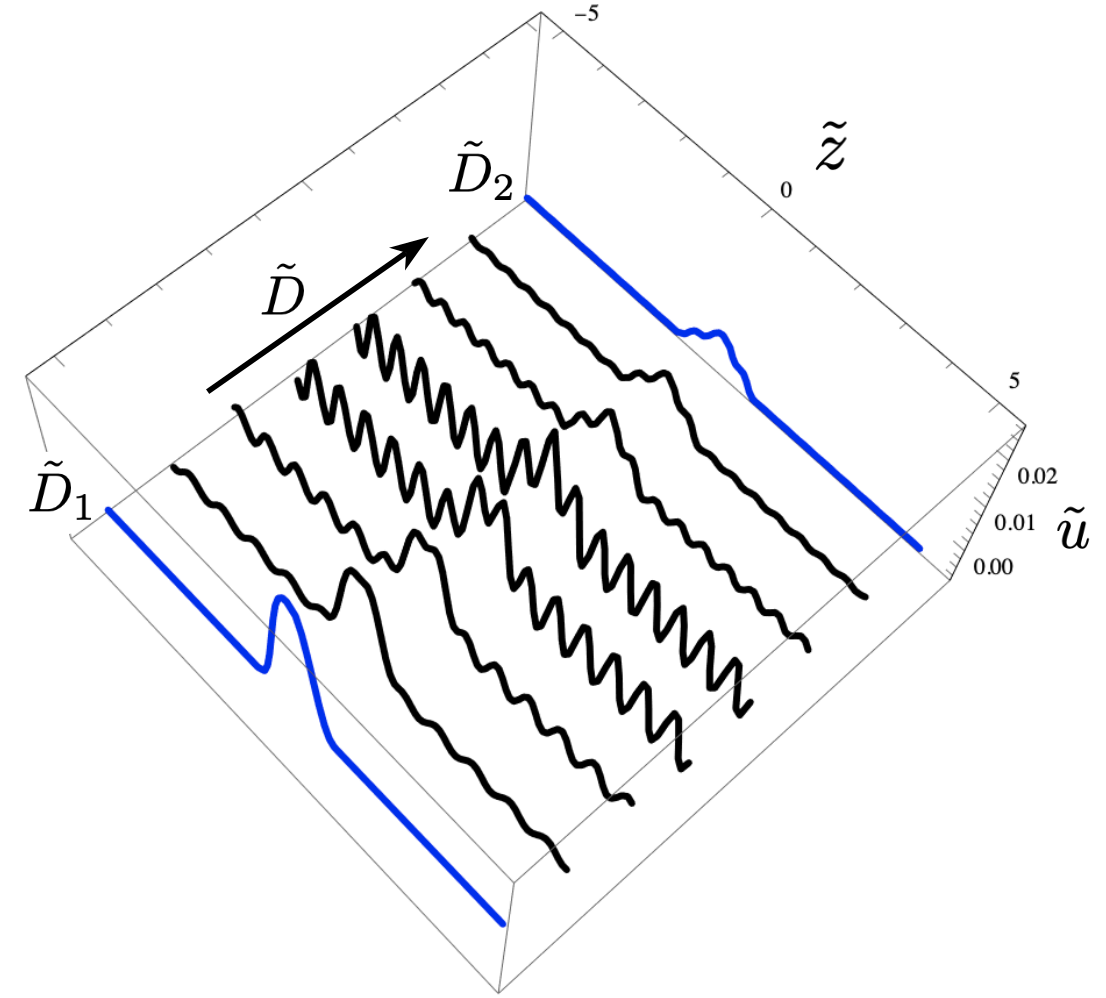} 
    \caption{\label{joydiv}\footnotesize{Solitary waves  with periodic tails (nanopterons shown in black) in the interval of driving velocities between $\tilde D_1=2 \pi / K $ and $\tilde D_2 = 4 \pi / K$. The corresponding compactons at  $\tilde D_1$ and $\tilde D_1$  are shown in blue. Here $K=1$, $E=0$.}}
\end{figure}

If we  now relax the requirement that the displacements vanish in the regions 1 and 4, and instead impose the weaker condition that   displacements remain bounded, we   need to impose  the constraints 
$ 
c_{12} = c_{42} = 0. 
$ 
If we also  require that  $c_{11}=c_{41} = 0$, so that the average displacements in the the regions 1 and 4 are equal to zero, we obtain  unique solutions of the system  \eqref{jumps1}, \eqref{jumps2}, and \eqref{jumps3}  for a continuous range of parameters $\beta>0$ (including compactons as   special cases) in the form 
\beq\label{nano14}
\tilde{u}_{1,4} (\tilde{z})= \frac{\cos\!\big((K \mp \tilde z)\,\beta\big)\,\tan\!\big(\tfrac{K\beta}{2}\big)}{\beta^{3}}
\eeq
in zones $1,4$,  and
\beq\label{nano23}
\tilde{u}_{2,3}(\tilde{z}) = \frac{K \beta - \tilde z\,\beta + \sin(\tilde z\,\beta) - \cos(\tilde z\,\beta)\,\tan\left(\frac{K \beta}{2}\right)}{\beta^3}
\eeq
in zones $2,3$, see a representative profile in  Fig.\ref{nano}.  Such semi-localized traveling waves with oscillatory tails reaching infinity are sometimes referred to as   \textit{nanoptera} or  {\it nanopterons} \cite{Beale1991, Boyd1991, MostonDuggan2023, Faver2019, Faver2021, Vainchtein2022,Kim2015, Widjaja2025}. The relation between nanopterons and compactons is illustrated in Fig. \ref{joydiv}.

In our case such solutions  are   characterized by nonzero energy radiation at $\pm \infty$ even though  under the assumptions (\ref{nano14}, \ref{nano23})the  energy loss or gain at the three jump discontinuities ($\mathcal{D}_{12}=\mathcal{D}_{23}=\mathcal{D}_{34}=0$) is equal to zero. To assess the  rate of energy exchange at infinity  we can  compute   the period-averaged energy flux. Using the definition 
\beq\label{averaging2}
\langle \mathcal{J} \rangle : = \frac{\omega}{2\pi}\int_0^{2\pi/\omega} \mathcal{J}(z)\,dz 
\eeq
where $\omega$ is the angular frequency of a periodic signal we obtain that   in our  oscillatory tails (regimes 1 and 4) 
\beq\label{averaging}
\langle \mathcal{J}_{1}\rangle =  \langle \mathcal{J}_{4}\rangle = \frac{c_0^2 D \rho (2\beta^2+\nu^2)}{2\beta^4}\tan\left(\frac{l \beta}{2 l_0}\right)^2. 
\eeq
One can see that these fluxes are  different from zero unless the parameter $\beta_n$ takes the discrete  values   given by \eqref{betan}.  Therefore, the traveling waves described by \eqref{nano14}–\eqref{nano23} are not exchanging energy  with $\pm \infty$)  only at quantized velocity values given by \label{Dn}. This also implies that only compact  pulses  are \emph{self-sustaining} in the above sense.

\begin{figure}[thb!]  
\centering
    \includegraphics[width=0.6\columnwidth]{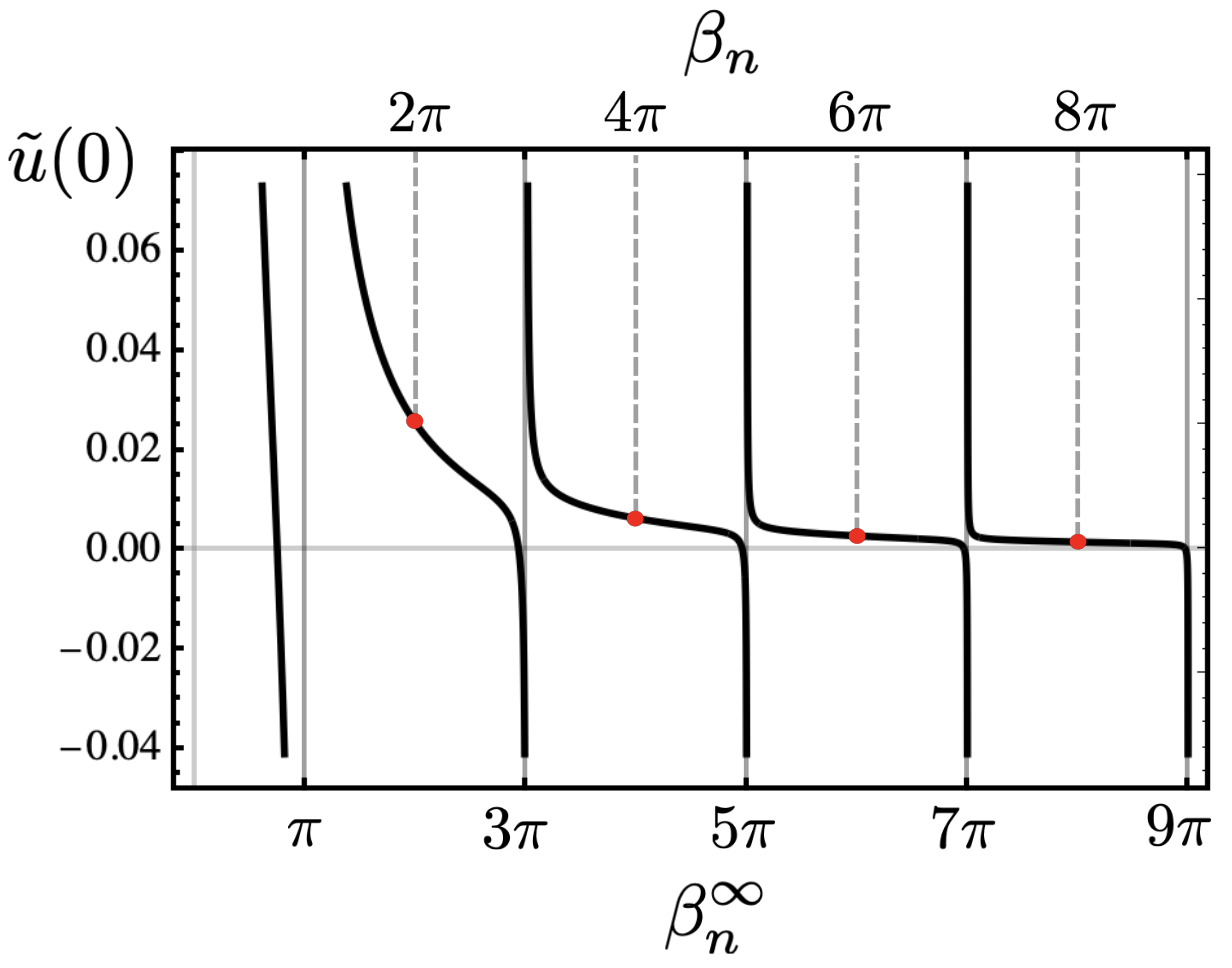} 
    \caption{\label{amplitude}\footnotesize{Amplitudes of the displacement field at $\tilde z=0$ as a function of the coefficient $\beta$. The vertical asymptotes (resonant regimes) correspond to $\beta=\beta^{\infty}_n$ (they are marked at the lower horizontal axis); the values $\beta=\beta_n$ corresponding to compactons  are marked on the upper horizontal axis. Here $K=1$.}} 
\end{figure}

Instead, as we show in  Fig. \ref{amplitude} the noncompact supersonic nanopterons exhibit  resonant behavior with diverging amplitude at  
\beq \label{averaging1}
\beta^{\infty}_n = \frac{l_0}{l} (\pi + 2 n \pi), \qquad n=0,1,2,... 
\eeq
This means, in particular,  that  in sharp contrast  with self-sustaining  compactons, the traveling pulses  corresponding to \eqref{averaging1} may exhibit arbitrary strong dependence on what is happening at infinity.  For instance,   the average fluxes in the tails \eqref{averaging} diverge for $\beta$ corresponding to \eqref{averaging1}.

To answer the question about the direction of the energy exchange with infinity in nanopterons, for instance, to decide whether it is dissipative or anti-dissipative,  we need to discuss the implied radiative energy transfer in more detail. 



 
  Note first that if we  average the oscillatory solutions in the tails   
\beq \label{averaging3}
\langle \mathcal{J} \rangle =  D_g  \langle w \rangle.
\eeq 
Here  we  introduced the   group velocity 
\beq
D_g= \frac{d\omega}{dk}
\eeq 
where $\omega(k)$ is the dispersion relation which is obtained substituting the ansatz $ 
u(x,t)= e^{i(kx-\omega t)}
$ 
into \eqref{main-equation-motion}. 
We obtain
$$
\omega^{2}=c_{0}^{2}\nu^{2}k^{2}+c_{0}^{2}l_{0}^{2}k^{4}
$$
which gives \beq
D_g=  \frac{c_{0} \!\left(\nu^{2}+2l_{0}^{2}k^{2}\right)}{ k \sqrt{\nu^2+l_0^2 k^2} }.
\eeq 
Observe next that in a traveling wave moving with velocity $D$, the emitted/received  ``effective phonons'' propagate with the phase velocity   
  \beq
 D_p= \frac{\omega}{k}=c_0 \sqrt{\nu^2+l_0^2 k^2}=D.
  \eeq 
One can see that 
\[
D<D_g<2D  
\]
with $D_g$ continuously interpolating between $D_g=2D$ in the bending-dominated case ($\nu=0$) and $D_g\rightarrow D$ in the extensional-dominated case ($\nu\rightarrow\infty$). In other words,  the presence of stretching elasticity lowers the group velocity but never below the  velocity of the pulse. Since, independently of the value of $\nu$, the group velocity $D_g$ remains strictly larger than $D$, the elastic energy is radiated    ahead of the moving pulse. Therefore nanoptera solutions with two sided oscillatory tails   will be able to  dissipate energy   at  $+\infty$  while also being able  to  receive energy from sources at   $-\infty$.

The developed understanding regarding the directionality of the radiative energy transfer can also shed light on the energetics behind the propagation  of the  quantized compactons. Note first   that, in contrast to what we have concluded for nanopterons,  for the steady propagation of the  compact mechanical pulses it is   necessary that around   the internal boundary $\tilde z=  K$,  where the activity is switched on,  energy is received, while    around the internal boundary $\tilde z=  -K$,  where the activity is turned off, energy is released. Moreover, since for compactons
\beq
\mathcal{J}(z)=D_n w(z)
\eeq 
 we can also conclude that  in regions 2, 3 so $\mathcal{J}(z)\geq 0$ while  $\mathcal{J}(z)= 0$ in regions 1,4.

  Furthermore, since $D_g >D$,  the energy   travels faster than the pulse itself and therefore there exists   a relative   unidirectional  energy transfer from the back of the pulse to its  front. More specifically,   the released mechanical energy at the rear   of the pulse travels to the front of the pulse and  the absorbed   energy at the front  is exactly equal to the released  energy at the rear. Effectively,  the  system is   harvesting the mechanical energy at one region and is relocating it with zero loss to another region. It is rather remarkable, that under this scenario,  no chemical energy is  needed to ensure the   mechanical  performance of the system.  In this sense,  while  the chemical  microscopic machinery  still expends energy  to generate the driving signal,  the compact  pulses  can be viewed as mechanically  autonomous.

\section{Subsonic regime}

\begin{figure}[h!]
\includegraphics[width=0.6\columnwidth]{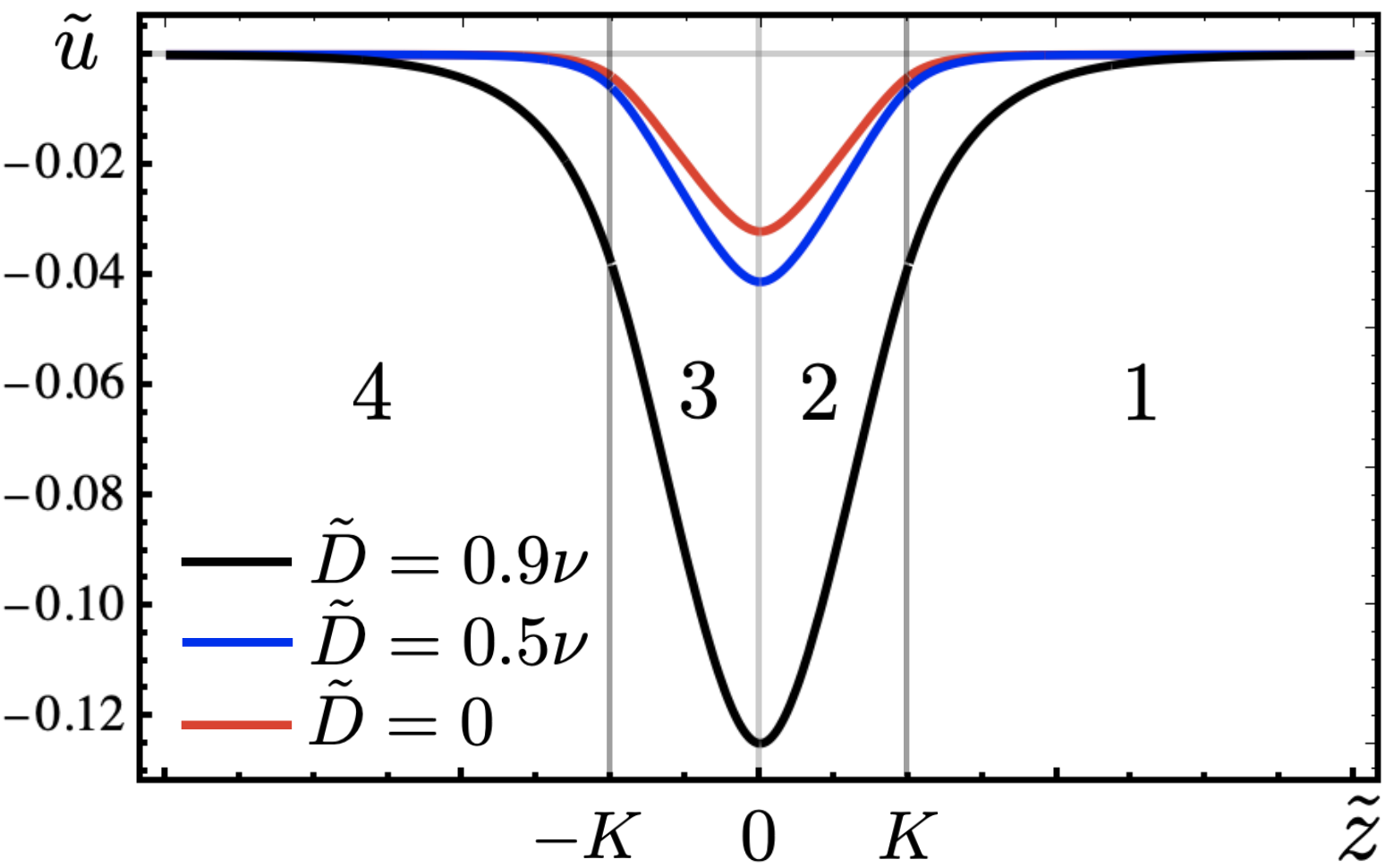}
\caption{\footnotesize{Dimensionless displacement field  in the subsonic regime with $\nu=5$ and different values of the velocity $\tilde D $.  }} \label{fig-exp-solitons}
\end{figure}

Here  we briefly discuss the  subsonic traveling wave solutions of \eqref{main} corresponding to
  \beq
\tilde D < \nu.
\eeq 
They require for their existence that either the stretching elasticity $E>0$ is sufficiently big or the  the velocity of the chemical signal  $\tilde D$ is sufficiently small. In such cases the  general  solution of the boundary value problem is the combination of linear and exponential functions, 
\beq\label{mainsolution-exp}
u_a(z) = b_{a1} + b_{a2} z + b_{a3}e^{\alpha z/l_0} + b_{a4}e^{-\alpha z/l_0}
\eeq
where  $a=1,..,4$ and where we have set
$$
\alpha=\sqrt{\nu^2-\tilde D^2}. 
$$ 
Under the assumptions
$$
b_{11}=b_{12}=b_{13} = b_{41}=b_{42}=b_{44} = 0
$$
the solution decays at $\pm \infty$    as desired. By imposing  the jump conditions \eqref{jumps1}, \eqref{jumps2}, we obtain a one parametric  family of the solutions.  Specifically, using $\alpha$ as the parameter, we obtain that in dimensionless variables  the solution in zones 1, 4  is
\beq \label{mainsolution-exp1}
\tilde u_{1/4}(\tilde z) =-\frac{(e^{K\alpha}-1)^2}{2\alpha^3}e^{-\alpha(K\pm\tilde z)}
\eeq
whereas in zones 2,3 it is 
\beq\label{mainsolution-exp2}
\tilde u_{2/3}(\tilde z) = \frac{e^{\mp\alpha \tilde z}-e^{-\alpha K}\cosh(\alpha\tilde z)}{\alpha^3} - \frac{K\mp\tilde z}{\alpha^2}. 
\eeq
In Fig.\ref{fig-exp-solitons} we illustrate the typical profiles (\ref{mainsolution-exp1},\ref{mainsolution-exp2})  corresponding to different values of $\tilde D$. In striking contrast to the  supersonic compactons shown in Fig. \ref{shape1} and characterized by the condition  $\tilde u\geq 0$, the compact subsonic pulses (\ref{mainsolution-exp1},\ref{mainsolution-exp2}) have instead $\tilde u\leq 0$.

This is related to the fact that while in the supersonic case we deal with anomalous negative effective stiffness  $E_{\text{eff}}< 0$, in the subsonic case  we are in a more conventional regime with $E_{\text{eff}}> 0$. To see the implication of this it is sufficient to observe that  at  $\tilde D=0$ the  static version of the solution (\ref{mainsolution-exp1},\ref{mainsolution-exp2}) can be interpreted as the classical buckling of a beam induced by a distributed pre-stress, see for instance \cite{paroni2015buckling,ren2023buckling}. 
In particular,  our static solution shows that in regions of active internal tension (zone 3, where $\sigma_a>0$), the system  exhibits  {\it shortening} $(\tilde u'<0)$, whereas in regions of active internal contraction (zone 2, where $\sigma_a<0$), it shows  {\it extention} $(\tilde u'>0)$. This behavior persists throughout the subsonic regime  but changes qualitatively in the supersonic regimes  where the  $\tilde u (\tilde z)$ profiles flip over due to the emergence of negative stretching elasticity. Note also that when the parameter  $\tilde D$ approaches the critical value  $\nu$  from any of the two sides  sides $\tilde D \lessgtr \nu$, which in dimensional variables means that the chemical signal reaches the critical velocity 
\beq \label{critical}
D =  c_0 \sqrt{\frac{E}{\sigma_0}}
\eeq
  the magnitude of the displacement field diverges. As many other critical regimes of this type, the limit \eqref{critical} may be of particular biological relevance, see for instance \cite{mora2011biological,munoz2018colloquium,hudspeth2024critical} 

Finally we mention that subsonic pulses do not interact energetically with $\pm \infty$ since the fluxes 
\beq
{\mathcal{J}}_{1/4}
= 
\frac{\varrho\, \tilde D\, c_0^3 \nu^2}{4(\tilde D^2 - \nu^2)^2}
\left(e^{K\sqrt{\nu^2 - \tilde D^2}}\right)^{4}
e^{-2(K \pm \tilde z)}
\eeq
 vanish  inside  the exponential tails. Also  the global dissipation is zero, due to the exact cancellation between the energy sink   at the dissipative singularity located at the leading edge of the active region  (the jump separating regions 1 and 2)  and energy source  located at the anti-dissipative singularity at the corresponding trailing edge (the jump separating regions 3 and 4) 
\beq
\mathcal{D}_{12}
= -\mathcal{D}_{34}
=
\frac{c_0^3 \tilde D \rho}{2(\nu^2 - \tilde D^2)}
\left(1 - e^{-K\sqrt{\nu^2 - \tilde D^2}}\right)^{2}.
\eeq
Given that $\mathcal{D}_{23} = 0$ which means that there are no energy sources or sinks  between regions 2 and 3,  we can conclude that $\mathcal{D} = \mathcal{D}_{12} + \mathcal{D}_{23} + \mathcal{D}_{34} = 0$ and that as long as the local sources and sinks are balanced, the corresponding mechanical system is energetically autonomous.

\section{Discussion}

 We have seen that the velocity of  compact  supersonic chemo-mechanical solitons, which are the only  relevant ones in the physically realistic regime when    stretching elasticity is negligible, can only take a discrete set of values.  Instead, a generic chemical driving would have produced  a highly dissipative mechanical response with energy escaping to infinity due to unavoidable generation of elastic radiation by the advancing mechanical pulse. The cancellation of such  radiative damping through a  particular choice of the chemical signal  can be achieved only due to  self organization   implying  an intricate   coupling between  chemistry and elasticity. In our highly schematic  model such coupling is characterized by a dimensionless quantity $K$ which brings together the parameters of the chemical activation ($l$, $\sigma_0$) and the elastic modulus $\kappa$. While in the context of biological systems, the   special regimes exhibiting compact pulses (whose stability is to be addressed separately),  can be viewed as a result of evolution, it is clear that artificial robotic systems of the same type can be directly controlled to exclude or at least minimize radiation  by parasitic elastic   waves.

\begin{figure}[h!]
\centering
\includegraphics[width=0.6\columnwidth]{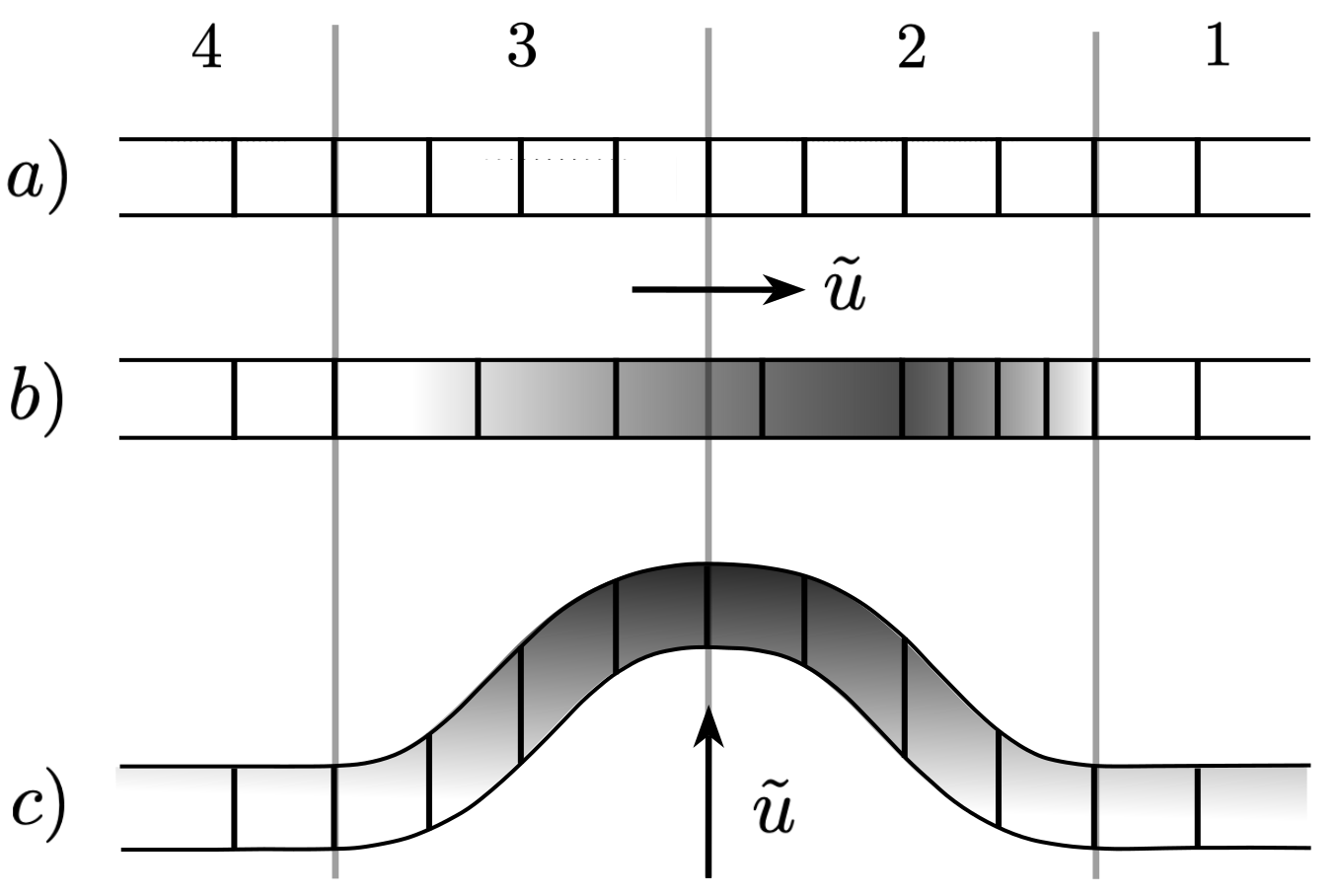}
\caption{\footnotesize{Schematic representation of a mechanical pulse advancing to the right. $a)$ Reference (undeformed) configuration; $b)$  deformed configuration in the case of longitudinal displacements; $c)$ deformed configuration  in the case of  transverse displacements.  The intensity of gray  is proportional to the magnitude of displacement.}} \label{fig:11}
\end{figure}
 
One possible  physical interpretation of the obtained quantized compact pulses can be in terms of a model of a peristaltic  ``pump''. To this end consider a  compacton configuration with the largest amplitude,  which corresponds to $n=1$. Then,  instead of viewing it as representing  a longitudinal deformation, driven by   localized tension and compression,   suppose that it represents a  transverse  deformation,  driven by active chemical pulse inducing a combination of positive and negative shear as it is schematically illustrated in Fig. \ref{fig:11}. In this case the resulting deformation takes the form of  a  localized ``bulge"   propagating along a  beam placed without cohesion on a flat surface. The beam need of course to be dynamically  pre-stressed by a chemical signal. The leading edge of the propagating bulge would then represent the  tip of an opening ``crack'',  while the trailing edge would indicate the location of its closure  side.  Interestingly, a concept of   such  self-healing   mechanical propagation of a localized  pulse  is used in non-biological fields as well fields from friction to  earthquakes \cite{perrin1995self,thogersen2021minimal}. 

Another relevant analogy would be with mechanical solitary waves in  nonlinear dispersive  passive  systems.   Since the corresponding localized wave packets are energy conserving, they   necessarily support an internal energy transfer between the leading and the  trailing segments. For instance, in the case of  discrete FPU-type   solitons,  one can interpret the leading edge as a partially developed  \emph{admissible}    shock wave  while  the trailing segment can be similarly viewed  as an   \emph{non-admissible}   shock wave \cite{dafermos2005hyperbolic}. While in such problems the nature of dispersion is different from our case  so that, in particular, the energy transport takes place from the front to  the back so that  the ``dissipative'' shock ``feeds'' (and in this way stabilizes) the ``anti-dissipative'' one  \cite{truskinovsky2014solitary,vainchtein2024solitary}.

\section{Conclusions}

In this paper we considered a   simple continuum model illustrating a possibility of chemically activated mechanical motion in biological systems. The crucial idea was to    represent  chemical degrees of freedom by  an external stress field. Our main result is the  development of  a prototypical model of a chemo-mechanical soliton driven by  a passive-to-active  transformation   at the front   and  a matching  active-to-passive  transformation    at the   rear. We showed that such solitons can be   generated by dynamic inhomogeneities  even  in the simplest   \emph{linear}    mechanical systems.  The obtained pulse solutions  can be interpreted as describing  spatially distributed chemo-mechanical   motors  converting   mechanically neutral   chemical    signals into   functional  mechanical activity. There is a considerable interest to reproduce  such biological mechanisms artificially in  soft robots  
\cite{agostinelli2018peristaltic,miller2020gait,pehlevan2016integrative,recho2024optimal}, and  in this sense the proposed   model  can serve   as a proof of concept.   Thus, while  the model predicts energy transfer with perfect efficiency one should have in mind that  we  neglected some sources of dissipation in the bulk including the ones associated with reaction-diffusion driving as well as  mechanical friction. Therefore future work   aimed at  fully quantitative designs should also account for  dissipation. The comprehensive study  must  also account for    the  functional work (cargo) and should  be  set   in a realistic 3D geometry.

\section {Acknowledgments} 

The authors thank P. Recho, A. Vainchtein, G. Mishuris and N. Gorbushin for helpful discussions. L.T. acknowledges the support   under the grants ANR-17-CE08-0047-02 and  ANR-21-CE08-MESOCRYSP.


\end{document}